\documentclass[acmtog, screen]{acmart}
\usepackage[false]{anonymous-acm}
\usepackage{tabularx}
\usepackage{multirow}
\AtBeginDocument{%
  \providecommand\BibTeX{{%
    \normalfont B\kern-0.5em{\scshape i\kern-0.25em b}\kern-0.8em\TeX}}}

\copyrightyear{2023} 
\acmYear{2023} 
\setcopyright{rightsretained} 
\acmConference[AIES '23]{AAAI/ACM Conference on AI, Ethics, and Society}{August 8--10, 2023}{Montréal, QC, Canada}
\acmBooktitle{AAAI/ACM Conference on AI, Ethics, and Society (AIES '23), August 8--10, 2023, Montréal, QC, Canada}
\acmDOI{10.1145/3600211.3604766}
\acmISBN{979-8-4007-0231-0/23/08}

\begin{document}

\title{Ground Truth Or Dare: Factors Affecting The Creation Of Medical Datasets For Training AI}

\author{Hubert D. Zając}
\authornote{Both authors contributed equally to this research.}
\email{hdz@di.ku.dk}
\orcid{0000-0003-0689-6912}
\author{Natalia R. Avlona}
\orcid{0000-0002-1009-1810}
\authornotemark[1]
\email{naav@di.ku.dk}
\affiliation{%
  \institution{University of Copenhagen}
  \streetaddress{Universitetsparken 1}
  \city{Copenhagen}
  \country{Denmark}
  \postcode{2100}
}

\author{Tariq O. Andersen}
\email{tariq@di.ku.dk}
\orcid{0000-0002-9342-5520}
\affiliation{%
  \institution{University of Copenhagen}
  \streetaddress{Universitetsparken 1}
  \city{Copenhagen}
  \country{Denmark}
  \postcode{2100}
}

\author{Finn Kensing}
\email{kensing@di.ku.dk}
\orcid{0000-0002-1392-5999}
\affiliation{%
  \institution{University of Copenhagen}
  \streetaddress{Universitetsparken 1}
  \city{Copenhagen}
  \country{Denmark}
  \postcode{2100}
}

\author{Irina Shklovski}
\email{ias@di.ku.dk}
\orcid{0000-0003-1874-0958}
\affiliation{%
  \institution{University of Copenhagen}
  \streetaddress{Universitetsparken 1}
  \city{Copenhagen}
  \country{Denmark}
  \postcode{2100}
}

\renewcommand{\shortauthors}{Zając & Avlona et al.}

\begin{abstract}

One of the core goals of responsible AI development is ensuring high-quality training datasets. Many researchers have pointed to the importance of the annotation step in the creation of high-quality data, but less attention has been paid to the work that enables data annotation. We define this work as the design of ground truth schema and explore the challenges involved in the creation of datasets in the medical domain even before any annotations are made. Based on extensive work in three health-tech organisations, we describe five external and internal factors that condition medical dataset creation processes. Three external factors include regulatory constraints, the context of creation and use, and commercial and operational pressures. These factors condition medical data collection and shape the ground truth schema design. Two internal factors include epistemic differences and limits of labelling. These directly shape the design of the ground truth schema. Discussions of what constitutes high-quality data need to pay attention to the factors that shape and constrain what is possible to be created, to ensure responsible AI design.                 

\end{abstract}





\begin{teaserfigure}
  \includegraphics[width=\textwidth]{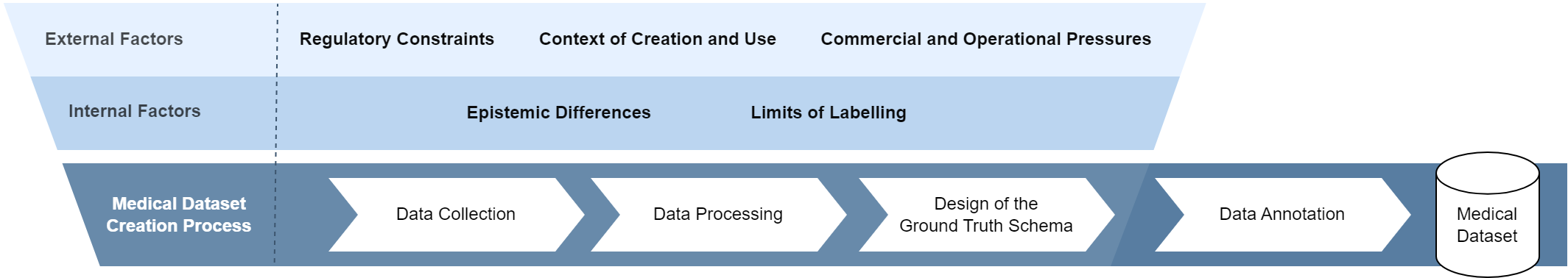}
  \caption{A simplified medical dataset creation process expanded with the design of ground truth schema and factors conditioning the pre-annotation stages.}
  \Description{A diagram depicting four stages of medical data creation: data collection, data processing, design of ground truth schema, and data labelling. Above the first three stages, there are depicted two levels of factors - external factors that include: regulatory constraints, the context of creation and use, and commercial and operational pressures, and internal factors that include epistemic differences, and limits of labelling.}
  \label{fig:teaser}
\end{teaserfigure}


\maketitle

\section{Introduction}
Advances in applications of Artificial Intelligence (AI) in the medical domain promise to improve efficiency, promote accuracy and bring cost savings across many areas of medical subspecialty, yet there are also many concerns about ethics and responsibility in the deployment of these technologies \cite{Dignum2020ResponsibilityIntelligence}. The idea of responsible AI has been extensively discussed in the literature and received much attention from both commercial entities and regulatory bodies \cite{Deshpande2022ResponsibleStakeholders, Mitchell2019ModelReporting}. There is considerable agreement that high-quality training data is key to the development of responsible AI systems \cite{HoltenMller2020ShiftingValue}. Yet research shows that the creation of high-quality data also tends to be an undervalued step in the development of machine learning systems \cite{Sambasivan2021EveryoneAI, Scheuerman2021DoDevelopment}. 

The process of dataset creation is typically broken down into three steps - data collection, data pre-processing and cleaning, and finally, data annotation\cite{Amershi2019, Nascimento2019UnderstandingSolutions}. This is especially so in the medical domain where high-quality training data is obtained through a range of annotation practices such as data quality enhancement \cite{Chen2021DataLearning}, generating labels using Natural Language Processing models \cite{Hallinan2022DetectionLabels}, deriving image labels from medical documentation \cite{Jain2021EffectInterpretation}, and following labelling guidelines and principles focusing on fairness and inclusion \cite{Leavy2021EthicalRace, Schumann2021AFairness}. This paper investigates the factors that affect the creation of high-quality medical datasets demonstrating that the preparatory work involved in the design of ground truth schema used in data annotation is an important preceding step that tends to be overlooked in the literature. Following the work of Mueller and colleagues \cite{Muller2021}, we define the ground truth schema as a collection of relational labels and metrics, as well as their definitions and examples that are used during data labelling.

Recent research on the creation of training datasets  \cite{Geiger2021GarbageData} has discussed annotation activities as a matter of power relations in projects crowdsourced in the Global South \cite{Miceli2022StudyingPower, Miceli2021DocumentingPractices, Miceli2022TheDispositif}, the social design of labelled data by domain experts \cite{Muller2021}, and annotation process recommendations \cite{Fort2016}. While understanding data annotation is important, data design work begins before the first data points are labelled. Data is always designed and constructed through situated and emergent processes \cite{Muller2021, Feinberg2017} as domain experts, data scientists, other stakeholders, and diverse political interests imprint their values on the data. However, little is known about the preparatory work necessary to produce high-quality data \cite{Hutchinson2021TowardsDatasets}. Accounts of decisions that shaped the datasets are rarely documented and get dismissed as soon as the data creation work concludes \cite{Scheuerman2021DoDevelopment},thus become impossible to inspect in the future \cite{Muller2021, Star1999LayersWork, Oakden-Rayner2019ExploringDatasets}. 

In this article, we consider \textbf{what factors affect the design of medical datasets prior to data annotation.} We ground our findings in ethnographic research conducted across three organisations developing medical AI for (I) screening chest x-rays, (II) supporting the diagnosis of lung and pancreatic diseases (III) automating patients-to-clinical trials matchmaking. We explore the decisions made by medical professionals, data scientists, designers, and other relevant stakeholders in their quest to create medical AI datasets in highly constrained environments. Our data include approximately 50 hours of observations, interviews with 46 medical professionals, data scientists, and designers, as well as observation notes, email communication, reports, and artefacts. We followed a grounded theory approach \cite{Charmaz2014ConstructingEd.}that led us to identify and define factors that influence the design of the ground truth schema that underpins the production of high-quality training data.

Our contribution is twofold. First, we identify five factors, three external and two internal, that influence medical dataset creation by affecting data collection, ground truth schema design, and data annotation stages (see Fig. \ref{fig:teaser}). The  external factors condition  the medical dataset creation processes by determining the data collection and shaping the possibilities for the design of ground truth schemas:
\begin{itemize}
    \item Regulatory Constraints 
    \item Context of Creation and Use
    \item Commercial and Operational Pressures
\end{itemize}
The internal factors define the negotiations between the medical and technical domains:
\begin{itemize}
    \item Epistemic Differences 
    \item Limits of Labelling
\end{itemize}

Second, we show how these factors affect the final shape and quality of the resulting medical datasets. While we define each factor separately for analytical purposes, the factors are interrelated and affect each other, structuring the limits of responsible data creation approaches. We argue that these factors condition the stages that precede data labelling and mediate the design of what is aspired to be responsible AI.

\section{Related Work}
While the idea of responsible AI has received much attention from both commercial entities and regulatory bodies, concerns about the quality of data and the challenges in the creation of quality data are increasingly in focus. The now-emerging guidelines list several data-related challenges as key obstacles that hinder the path towards responsible AI: skewed data (issues that originate during data collection), tainted data (issues that stem from labelling e.g. hidden stratification \cite{Oakden-Rayner2020HiddenImaging}), or limited features (an inadequate number of features represented in data) \cite{BarredoArrieta2020ExplainableAI}. There is broad agreement that dataset creation processes deserve greater attention, despite scholars repeatedly pointing to a strong bias against data work \cite{Denton2021WhoseAnnotation, Sambasivan2021EveryoneAI, Scheuerman2021DoDevelopment}.

\subsection{How datasets are created and annotated}
In computer science, dataset creation is often seen as an activity constituting a step in the larger development processes of ML-based systems \cite{Nascimento2019UnderstandingSolutions, Hill2016TrialsStudy, Chen2019HowHealthcare, Wang2019Human-AIAI, Amershi2019}. However, scholars have also discussed the dataset creation process on its own merits. For example, Hutchinson drew parallels between software development and dataset creation practices by sharing conceptual stages like requirement analysis, design, implementation, testing, and maintenance\cite{Hutchinson2021TowardsDatasets}. Similarly, increased focus can be observed in the medical area, where researchers describe in greater detail the creation of publicly available medical datasets \cite{Bustos2020, Irvin2019, Johnson2019, Wang2019, Mendonca2013PH2Benchmarking, Demner-Fushman2016}. Typically, dataset creation is described as a process that spans all activities related to work on medical data, collected under the umbrella of data collection, data cleaning and processing, and data annotation.

Data annotation is one of the most researched aspects of dataset creation. Data annotation or labelling usually happens as part of the curation or preparation step of larger data science projects, following data acquisition and cleaning, and preceding feature engineering \cite{Amershi2019}. These activities are usually iterative and highly collaborative. Linguistic scholars and Natural Language Processing researchers \cite{Fort2016, Voormann2008AgileCreation, Hovy2010TowardsLinguistics} offer guidance on how to carry out data labelling. They distinguish three focal points: the creation and improvement of an annotation guide \cite{Fort2016}, schema \cite{Voormann2008AgileCreation}, or manual \cite{Hovy2010TowardsLinguistics}; the labelling performed by trained annotators; and the adjudication of the annotated data.

In this paper, we use the terms data labelling and data annotation interchangeably and understand them as the action of assigning and adjudicating predefined labels to concrete data points. When considering this step alone, there is a multitude of decisions that need to be taken to complete it. Scholars have pointed to data annotation activities as a site of political struggle, challenges to the labour conditions, as well as the stage in dataset creation that can result in adverse downstream outcomes for trained models. For example, Schumann et al. \cite{Schumann2021AFairness} and Hanley et al. \cite{Hanley2021ComputerText} demonstrate how the design of categories (or labels) can reinforce harmful stereotypes and exclude underrepresented groups of people. Badly annotated data can reduce the performance of AI models \cite{Reidsma2008ReliabilityLimits, Chen2021DataLearning, Mitchell2019ModelReporting, Hallinan2022DetectionLabels, Jain2021EffectInterpretation} and perpetuate exclusion and inequality \cite{Leavy2021EthicalRace, Schumann2021AFairness}.
 
In the medical domain, data annotation challenges can be compounded by the requirements for specialised knowledge and training. Despite initiatives like the Unified Medical System \cite{Lindberg1993TheSystem}, the clinical meaning of labels can be unclear \cite{Oakden-Rayner2019ExploringDatasets}, and medical knowledge remains difficult to capture for computer use. Li and colleagues \cite{Li2022Inter-System} explored the inter- and intra-rater agreement between six radiologists of different experience levels when labelling chest x-rays. In some cases, even the experienced radiologists reached only a moderate level of agreement with themselves \cite{McHugh2012InterraterStatistic}. This could occur due to not following the best medical practices when labelling data, due to resource constraints \cite{Sambasivan2021EveryoneAI} or because of the disconnect between the practices of labelling and the actual usage of medical data in regular practice \cite{Oakden-Rayner2019ExploringDatasets}.

What much of this research points to is the fact that labelling and annotation as practices are heavily reliant on the creation of annotation guides and schemas \cite{Fort2016}. Yet, despite the growing interest in the creation of datasets, current discussions tend to omit and overlook the pre-labelling activities and their potential impact on the quality of the resulting training data \cite{Zajac2022DesigningScientists}. 

\subsection{The design of the ground truth schema}
Many scholars investigated the dynamic and situated work of domain experts, data scientists, designers, and other stakeholders engaged with data \cite{Seidelin, Henriksen2020BuildingHealthcare, Muller2019HowData, Sambasivan2021EveryoneAI}. For example, Muller and colleagues investigated how domain experts label data, highlighting that the ground truth contained in datasets is a human contribution resulting from improvised and iterative adjustments to principled design processes \cite{Muller2021}. Discussing the design of ground truth schema implies that ground truth captured in medical AI datasets is not an objective representation of reality but is a result of a situated design process \cite{Bowker2000SortingConsequences}. In other words, data is never raw \cite{Gitelman2013RawOxymoron}, instead, all data is actively constructed \cite{Aroyo2015TruthAnnotation, Pine2015TheAction, Miceli2020BetweenImposition}. Feinberg emphasises the importance of recognising the subjectivity involved in dataset creation and the need to consider the potential biases and limitations inherent in choices that stem from the social and organisational context in which data is produced \cite{Feinberg2017}.

Researchers who investigate AI datasets suggest that access to all of the ``subtle design decisions'', made during the dataset creation, is vital to ensuring a high-quality labelling process \cite{Oakden-Rayner2019ExploringDatasets, Fabris2022AlgorithmicFar} and thus high-quality datasets. However, documenting design decisions in data science work is not common \cite{Pine2015TheAction, Zhang2020HowTools, Rule2018ExplorationNotebooks}. To address this gap, researchers developed a range of documentation frameworks to support the accountability, use, and maintenance of complex datasets \cite{Anik2021Data-CentricTransparency, Miceli2022DocumentingWork}. These frameworks range from general purpose and qualitative - Datasheets for Datasets \cite{Gebru2021DatasheetsDatasets}, NLP-focused - Data Statements \cite{Bender2018DataScience}, quantitative - Dataset Nutrition Label \cite{Holland2018TheStandards}, to fairness focused - data briefs \cite{Fabris2022AlgorithmicFar} and accountability \cite{Hutchinson2021TowardsDatasets}. Some of these tools \cite{Gebru2021DatasheetsDatasets, Fabris2022AlgorithmicFar, Hutchinson2021TowardsDatasets} include a query for the origin of the labels, but most do not pay much attention to the pre-labelling activities involved in annotation schema creation.

While the existing scholarship has problematised the stage of the data labelling and the power relations and conditions affecting the data annotation work \cite{Muller2021}, little is known about the stages preceding the data labelling. Particularly, how these stages influence the final shape of medical datasets. We explore the collaborative and situated work of medical professionals, data scientists, and designers that takes place before the labelling stage, within the design stage proposed by Hutchinson et al. \cite{Hutchinson2021TowardsDatasets} or the preparatory work proposed by Fort \cite{Fort2016}.

\section{Methodology}
We investigated three organisations in the Global North developing medical AI-based systems that engaged in the medical dataset creation processes. We focused on the work conducted before the data annotation task by participants described in Table \ref{tab:participants}).

{\footnotesize

\begin{table*}[]
\begin{tabular}{lll}
\toprule
\textbf{ORG I} & \textbf{Position}    & \textbf{Exp.} \\
\midrule
P1        & Radiologist          & Junior    \\
P2        & ML Engineer          & Senior    \\
P3        & ML Engineer          & Senior    \\
P4        & Computer   Scientist & Senior    \\
P5        & Data Scientist       & Senior    \\
P6        & Radiologist          & Senior    \\
P7        & Radiologist          & Senior    \\
P8        & HCI Researcher       & Junior    \\
P9        & HCI Researcher       & Senior   \\
&&\\
&&\\
&&\\
&&\\
\bottomrule
\end{tabular}
\hfill
\begin{tabular}{lll}
\toprule
\textbf{ORG I} & \textbf{Position}    & \textbf{Exp.} \\
\midrule
P10       & Radiologist       & Senior    \\
P11       & Radiologist       & Junior    \\
P12       & Radiologist       & Junior    \\
P13       & Radiologist       & Mid       \\
P14       & Radiologist       & Senior    \\
P15       & Radiologist       & Senior    \\
P16       & Radiologist       & Senior    \\
P17       & Radiologist       & Senior    \\
P18       & Radiologist       & Senior    \\
P19       & Radiologist       & Senior    \\
P20       & Radiologist       & Senior    \\
P21       & Physician         & Junior   \\
&&\\
\bottomrule
\end{tabular}
\hfill
\begin{tabular}{lll}
\toprule
\textbf{ORG II} & \textbf{Position} & \textbf{Exp.} \\
\midrule
P21      & Data scientist       & Senior        \\
P22      & Product Owner        & Mid           \\
P23      & Strategic Designer   & Senior        \\
P24      & Data scientist       & Mid           \\
P25      & Usability Designer   & Senior        \\
P26      & Data scientist       & Senior        \\
P27      & Data scientist       & Senior        \\
P28      & Data Designer        & Mid           \\
P29      & Interaction Designer & Senior        \\
P30      & Data scientist       & Senior        \\
P31      & Data Designer        & Senior        \\
P32      & HCI Researcher       & Mid           \\
P33      & Data Designer        & Senior   \\
\bottomrule
\end{tabular}
\hfill
\begin{tabular}{lll}
\toprule
\textbf{ORG III} & \textbf{Position}   & \textbf{Exp.} \\
\midrule
P34     & Product owner      & Mid           \\
P35     & Software Engineer  & Junior        \\
P36     & Software Engineer  & Mid           \\
P37     & Software Engineer  & Mid           \\
P38     & Data Scientist     & Mid           \\
P39     & Data Scientist     & Senior        \\
P40     & UX Designer        & Senior        \\
P41     & Software Developer & Mid           \\
P42     & Medical Operations & Senior        \\
P43     & Quality Assurance  & Senior        \\
P44     & UX Designer        & Mid           \\
P45     & Neurobiologist     & Senior        \\
P46     & Product Owner      & Mid  \\
\bottomrule
\end{tabular}
\caption{List of participants, their simplified positions, and experience levels. Respectively in ORG I (working group), ORG I (feedback group, participants 10-14 were located in the northern European country, and participants 15-21 were located in the East African country), ORG II, and ORG III.}
\end{table*}
\label{tab:participants}
}

\subsection{Research context and data collection}

\subsubsection{ORG I} was an interdisciplinary collaboration between academia, business, and the public healthcare sector, aiming to create AI-based chest x-ray prioritisation software for global use. The project's first step was designing the ground truth schema for labelling chest x-rays, which is the process investigated in this study. 

Our engagement in ORG I spanned May 2021 to Feb 2023. During that time, we conducted participatory observations of the design process of the ground truth schema. The working group developing the system was based in a Northern European country (Table \ref{tab:participants}.1). Additionally, a feedback group comprising medical professionals from the Northern European country and an East African country provided feedback on the schema (Table \ref{tab:participants}.2). We participated in fifteen working group meetings ranging from 26 minutes to 2 hours and 12 minutes in length.  Additionally, we conducted twelve interviews and observed external medical professionals evaluating and providing feedback on the intermediate results of the design work. Additional material included observation notes, meeting summaries from other participants, a work progress report, email communication, and produced artefacts - a labelling guide and the ground truth schema.

\subsubsection{ORG II} was a large tech company  in Western Europe with part of the business involved in the development of complex medical devices. We primarily engaged with sections of the company that focused on the development of AI-based diagnostic tools and systems for oncological radiology.

Our work with ORG II was split into a preliminary exploratory period online from February to May 2022 and in situ participant observations and semi-structured interviews conducted in June 2022 in a Western European country. Due to the size of the organisation, we employed snowball sampling. In ORG II, we conducted thirteen semi-structured interviews with experts (Table \ref{tab:participants}.3), with an average duration of 65 minutes.

\subsubsection{ORG III} was a mid-size start-up in Western Europe that aimed at developing an AI-based platform for matching patients with advanced clinical trials for new drug and experimental procedure development. The company primarily dealt with two data sources. First, they collected data from medical practitioners and their patients. Second, they collected data from public registries in the EU and US and pharmaceutical companies about clinical trial requirements or experimental treatments. Their goal was to match the patients with unmet medical needs and their physicians with the requirements of BioPharma companies that need to enhance drug development and recruit participants for clinical trials.

Our engagement with ORG III spanned February to May 2022. The preliminary period involved online semi-formal meetings and interviews from February to April 2022. In situethnographic research was conducted during May and June 2022 at the headquarters of ORG III in Western Europe. We conducted participant observation by joining the daily stand-up sessions of the engineering department and shadowing the workflow of the AI team experts leading the data labelling process for the match-making platform. In total, we interviewed 13 participants (Table \ref{tab:participants}.4).

\subsection{Data analysis}
The main focus of our analysis was to identify factors affecting medical dataset creation. We analysed decisions made during the design work, tensions and misunderstandings that needed to be reconciled, looking both outside and within the organisations where the design work took place. We explicitly decided to explore the wider socioeconomic factors that condition the medical dataset creation and influence the final AI-based systems even before the first label is annotated. 

Data analysis relied on techniques of grounded theory and situational analysis \cite{Charmaz2014ConstructingEd., Clarke2005SituationalAnalysis}. First, we conducted line-to-line open coding, coming up with 850 initial codes. We then reflexively proceeded to thematic coding, in an iterative manner, discussing the themes and patterns emerging in our three sites of ethnographic inquiry. During this step, we designed visual maps to lay out the human, technological, and discursive dynamics of the organisations under study \cite{Clarke2005SituationalAnalysis}. Second, we conducted axial coding to reflexively group the available themes into dimensions. Finally, we assessed these dimensions against the codes and situational maps, converging on the five final factors (regulatory constraints, context of creation and use, commercial and operational pressures, epistemic differences, and limits of labelling).

\subsection{Positionality statement} Our qualitative data was obtained from three health-tech organisations in the Global North. The analysis was shaped by the following standpoints. First, we differentiated our roles in studying the three organisations. Researchers in ORG I had the dual position of the expert who on the one hand designed the labelling software, whilst they conducted participant observation and semi-constructed interviews in order to study the process of the ground truth schema design. Researchers working with ORG II and ORG III employed ethnographic methods as a research approach without having a prior engagement with the organisations. Second, we are researchers currently working for Northern European institutions. Third, we have mixed epistemic backgrounds in computer science and law and policy. Finally, we emphasise the situatedness of our research, which focuses on the development of medical AI at the specific loci of our studied organisations. We acknowledge that the factors we identify as defining  the medical dataset creation bear the geographical and epistemic limitations of the Northern European context. On this note, we acknowledge that the divide between Global North and Global South we make below has been problematised by scholars in human geography and decolonial studies as a limiting one, reinforcing stereotypes and reducing the polyphony of southern standpoints \cite{Waisbich2021BeyondPolyphonies, Horner2020TowardsDevelopment}. For this reason, we use this divide in this paper to (I) acknowledge the limitations of our standpoints in a northern institution and the privilege of our funded projects; (II) tackle assumptions about data universalism \cite{Milan2019BigUniversalism} by showing the particularities of the northern context in medical datasets creation and their effect on the intended use of such data in different contexts.

\section{Findings: five factors that influence medical dataset creation}
The datasets used for medical AI benefit from the impression that they are a result of an age-old medical practice that is seamlessly transitioning to the digital age, unaffected by external influences, and focused on the pursuit of medical excellence. However, the reality is often different. Our ethnographic data suggest that even before medical professionals have had the chance to annotate or make their first label, many critical design decisions have been made, which frame the labelling space, thus limiting the extent to which medical professionals can use their expertise. 

Our analysis challenged our initial understanding of the dataset creation process drawn from the literature. Our data made clear that the preparatory work should be conceptualised as a crucial stage in dataset creation taking place before data labelling because it defines what becomes captured as ground truth within a training dataset. This is the step where the ground truth schema is designed, which, when applied to an unlabelled dataset through expert annotation, embeds the intended ground truth within it. 

We identified five factors that influenced the creation of medical datasets in the organisations we studied. Three of these factors were external to the activities directly involved in pre-labelling activities. External factors defined and delineated the limits and possibilities for labelling activities. Two internal factors on the other hand affected the negotiations around what needed to be labelled and how the labelling was to proceed through the design of the schema. Below we describe each factor and demonstrate how they affected the final shape of the medical datasets focusing on the data collection and ground truth schema design stages.

It is important to note that the organisations and processes examined in this paper were largely driven by data scientists as the owners of the dataset creation process, with representatives of other domains contributing to the dataset creation activities. As a result, data science as an epistemology dominated the design work by primarily embedding data scientists' perspectives, inadvertently compromising other domain-based practices and understandings. As datasets in our research were created for the purpose of AI development, the power distribution was uneven, leaving little room for misconceptions from data scientists to be challenged and addressed.

\subsection{External factors: defining the ground truth schema design space}
Despite the best intentions of the experts engaged in the medical dataset creation process, many of their decisions and actions were structured by different external factors. We identified three such factors - \textbf{Regulatory Constraints,  Context of Creation and Use}, and \textbf{Commercial and Operational Pressures} - that shaped the space of medical dataset creation and thus influenced the final shape of the datasets themselves even before the labelling could begin (Table \ref{tab:external}). Each factor consists of several distinct features. We describe these below in detail. 

{\small
\begin{table}

\begin{tabular}{p{\linewidth}}
\toprule

\textbf{Regulatory Constraints}
\begin{itemize}
    \item Extent of Collected Data
    \item Predetermination of Purpose
\end{itemize}

\textbf{Context of Creation and Use}
\begin{itemize}
    \item Geographic context of use
    \item Demographic context of production
    \item Linguistic context
\end{itemize}

\textbf{Commercial and Operational Pressures}
\begin{itemize}
    \item Business model and organisation scalability
    \item Competition and health tech market
    \item Intended future use within healthcare type
\end{itemize}\\
\bottomrule
\end{tabular}
\caption{External factors and their dimensions}
\label{tab:external}
\end{table}
}

\subsubsection{Regulatory constraints}
The medical data space is highly controlled through a variety of local, national, and international regulatory constraints. This was particularly challenging for the data collection step of the process. We observed two areas where compliance with regulatory standards affected the creation of medical data: \textbf{the extent of the collected data} and \textbf{the predetermination of purpose}. Experts in all of the organisations we studied were concerned about compliance with diverse standards that intersected with their work on medical dataset creation. These standards originated from European binding legislative acts, international standard organisations, or industry standards. GDPR, the main legal standard for data protection in the European Union, was the most prominent example of a binding legislative act, regulating the conditions under which personal data is collected and processed. The industry and international organisations imposed, among others, ISO 2700013001, HIPAA, and Good Medical or Good Manufacturing Practices. In ORG III, a data scientist (P39) listed 21 unique regulations they felt they needed to consider. As a larger and more mature organisation, ORG II also had internal ethics boards, which at times imposed even stricter interpretations. However, these standards and limits legitimised the data collection and processing activities. 

\textbf{Constraints on data collection.} While experts in all organisations were striving to create what they saw as high-quality data, complying with relevant regulatory standards required concessions from all participants. For data scientists, the regulatory constraints delimited what data was available for collection, at times inadvertently introducing bias in different ways. For example, P26, a data scientist from ORG II, explained: \textit{"what is the data that we are allowed to use, especially if you look at ... bias ... people will want to look at bias and, and see if ... their product was fair to all, some demographics, and [we are] just not able to use the data because of privacy issues or GDPR"}. Similarly, in ORG II, the contractual agreement with a single local hospital, on the one hand, provided a controlled supply of high-quality data, on the other hand, reduced data representativeness: \textit{"we have a strong relationship with them. How do you expect that the data is not going to be biased right?"} (P24). While ORG II was able to create highly detailed and structured training data for their models, this data was clearly not representative of populations that would eventually encounter the resulting technologies.

Limitations imposed on data collection could compromise the resulting datasets in ways that created challenges for subsequent data creation steps. For example, participants of ORG I could collect only chest x-rays and their linked radiological reports.  Privacy concerns here also resulted in the loss of the chronological links between the images during data collection. This selection significantly diverged from the usual assortment of data available to radiologists in clinical practice, introducing challenges at the later stages of medical dataset creation, such as schema creation and annotation.

\textbf{Regulatory standards and contractual agreements determined the purpose and context of use}. Data protection regulations have recently focused intently on the purpose of use as one area of emphasis, tied to notions of data minimisation and data subject notification. Companies in our research had to negotiate the legal basis for their data collection with contracted data providers such as hospitals. For example, GDPR and contractual agreements with a local hospital bounded ORG II to use the collected data within the predefined purpose and context. Deviations from the initially stated purposes and context of use required new agreements that could be obtained only through significant time and resource investments. As a product owner (P22) explained the process of collecting data from the local hospital, \textit{"maybe the new study that we want to do has a slightly different scope and it's not covered by the original contract, then we need to make a new contract"}. ORG I encountered a similar predicament where the data collection phase was negotiated based on what the data scientists believed to be a necessary and sufficient dataset given the available resources and legal constraints of local regulations. By the time domain experts explained that the dataset was lacking important data dimensions, it was too late.

\subsubsection{Context of creation and use}
The context of production and the context of use influenced the creation of medical datasets. In our studies, each medical dataset was created for a specific intended use that was embedded in the collected medical data, e.g., clinical trial repositories, hospitals, and patients. These sources cover specific geographical populations, which has consequences for the final medical dataset. We identified three dimensions where that influence was prevalent: \textbf{the geographic context of use, the demographic context of production}, and \textbf{the linguistic context} (Table \ref{tab:external}).

\textbf{The geographic context of use affected the selection of labels.} While medicine strives to deliver replicable results that generalise across populations, the ground truth schemas are designed to serve specific needs in specific contexts. Some of them are defined by the intended use of the future AI-based systems in the geographic context, in which they are going to be used. In ORG I, the project group designed the first version of the ground truth schema based on local data from a Northern European country. As a result, the first version of the schema captured the locally prevalent conditions well but missed conditions relevant within the countries of intended use, which were almost never encountered locally. To account for that, direct and indirect input from medical professionals from the East African country was collected and incorporated into the schema during joint design work, as seen in this exchange between a radiologist and an ML engineer. \\ 
\textit{"So if you wanted that in the hierarchy, it could be there."} (P1) \\
\textit{Is it aortic unfolding? Because I clearly remember this sentence from [the East African country] reports, "aortic unfolding due to chronic hypertension}" (P2). Yet despite having a broader ground truth schema, the same project also struggled to ensure enough examples of common medical conditions across expected countries of use available for annotation, since the data was originally only collected from one country. 

\textbf{The demographic context affected representativeness concerns} In both ORG I and ORG II, data in medical datasets were collected from a single country, which had several consequences. For example in ORG II, the data was predominantly collected from a single local hospital, where ORG II had a contractual agreement. Not only was this problematic due to a more homogeneous patient population, but the collected medical imaging data originated on machines from the same producer. This created many concerns since imaging machines from different manufacturers often produce slightly different artefacts in their output. Yet the information about which machines were used to produce the images was rarely included in the resulting dataset. 

Similarly, due to the characteristics of the population embedded in medical datasets, experts worried about how portable the resulting AI models would be. As a usability designer (P25) from ORG II noted, \textit{"you can have all sorts of differences in patient demographics ... and you cannot just apply a model that you train on population A to population B"}. However, despite the designers' and data scientists' awareness, a senior radiologist from the East African country emphasised that \textit{"in the [developing world]\footnote{edited to avoid pejorative language} we are usually consumers, not producers of tech. We may find ourselves hitched to tech that doesn't serve our needs"} (P15). When evaluating the ground truth schema, the same medical professional elaborated, \textit{"I've done this for 10 years since my graduation. I've never seen certain diseases like cystic fibrosis, but whenever I read the books, there's a lot of stuff about cystic fibrosis [prevalent in the Global North],"} which highlights the effect of local ground truth schemas on the transferability of the final AI-based systems.

\textbf{Linguistic context and local understanding of medical terms challenged the application and transferability of the ground truth schemas.} The design of ground truth schemas included naming the labels, defining and organising their relations, and providing examples. However, medical concepts are not always used in the same way across different countries. In ORG I when discussing the naming convention for a chest x-ray finding, one radiologist noted \textit{"I know that it's not proper, but [in the Northern European country] they use \textit{'infiltrat'} as a synonym of consolidation ... I think the direct translation consolidation would be \textit{'consolidering'} but they don't use that, they use \textit{'infiltrat'}... I think maybe our infiltrate is broader"} (P1). As a result, a presentation of infiltration by an AI-based system could be understood differently by medical professionals from different countries. To account for that, data scientists and medical professionals evaluated the ground truth schema against English translations. In ORG III, which operates globally, the data scientists and designers recounted a similar challenge of re-translating medical terms during the data annotation process. The limitations of the locality of medical terms prohibited the aspiration of designing a ground truth schema that can operate universally. As a UX designer (P40) remarked: \textit{"there are also challenges around that because different cultures will refer to different diseases in different ways. It's global and we re-translate some of our stuff into different pages. We also have to consider localisation, how you turn this medical term into a layman term, but that's also relevant in like different countries as well."}

\subsubsection{Commercial and operational pressures} 
The three organisations each had a different business model and exhibited different relations to the market and the public sector. This often determined the availability of the resources (human and material) allocated for dataset creation and affected the organisations' ability to collect data and design the ground truth schema. We identified three dimensions of commercial and operational pressures (Table \ref{tab:external}): \textbf{business model and scalability of the organisation, the competition in the health tech market}, and \textbf{intended future use within the healthcare type}. 
 
\textbf{The business model and scalability of the organisation determined the amount of collected and labelled data.} Every investigated organisation represented a different business model. ORG I intersected with the public sector, whilst ORG II and III were situated entirely in the private sector. The business models of the organisation determined the way in which data was collected. The business model of ORG III relied on providing free use of the AI-based platform to patients but also providing paid services to BioPharma by enrolling patients into clinical trials. To do that, ORG III collected data from the public clinical trial registries in the EU and US, as well as patient medical information. Such data collection was heavily dependent on the organisation's scalability, as well as the "fine" balance between the data requested by their BioPharma clients and the data that could have been collected. As a data scientist (P38) explained: "sometimes it's difficult to decide what kind of data you collect, right? Or what patients. (...) there's a balance between what's actually feasible to collect and what will give us the highest chance of getting as much data as possible. So those I think are tricky decisions." These conditions affected how much data was finally collected, hence, the ideal of representativeness of the created dataset was compromised.

In ORG I, the budget allocation for the data annotation process played a vital role in the amount of data possible to be labelled by medical professionals. Due to the high cost of labelling by experienced medical professionals, ORG I had to cap the maximum number of labelled images. This cap limited the number of distinct labels that could be annotated in the created dataset and remained statistically significant. "We have a limited budget for the test data that we can collect because we need several radiologists board-certified possibly to look at images" (P3). The limited resources defined the amount of data that was possible to be annotated, putting ORG I at a competitive disadvantage:  "What the [competitors] do (...) there is no way we can reach what they do. They have 127 findings and they use a hundred plus radiologists to annotate, and they annotated 800,000 images each image by three radiologists. So the scale is completely different" (P2).

\textbf{Market standards and industry competition affect the design of the ground truth schemas.} Since all organisations under study operated in the health tech sector, the experts engaged in the processes of designing ground truth schemas had to both consider existing state-of-the-art solutions and methods, as well as address market competition. In ORG I, the choice of a specific machine learning model architecture was dictated by the industry standard. However, this choice had consequences for the label needs during the design of the ground truth schemas. At the same time, addressing market competition influenced the work on the ground truth schema design, as seen here, "so this is [a competitor's system] and this is their output. they ... split consolidation and nodules, which at this stage of the hierarchy we are not doing. And so I was wondering why we're not doing it" (P2). In this organisation competition directly influenced the design work.  

Due to the large size of ORG II, the matter of competition fed to internal business processes whose results other experts relied on during the dataset creation, as explained by a product owner (P22), "it's a combination of ... alignment with the business priorities and that is also strongly driven by customer requests and customer demands. So that is actually very important ... try to find the alignment". Finally, market competition created time pressures that could structure and limit how data creation had to be organised: "if you want to validate something properly, it costs time. If you want to validate across domains, it costs time. And we are often in very competitive domains where being fast to market or, or fast at the FDA is also important. So there are some time trade-offs, need to be made there." (P27). 

\textbf{The intended use and type of healthcare system affected the content and the level of detail of the ground truth schemas.}
Visions of future intended use permeated the design work on the ground truth schemas. The imagined intended use of a future AI-based system factored into decisions about the validity of label choices. Imagined use did not fit in with current domain-specific practices and resulted in confusion and concerns during the design of the ground truth schema. Consider the following discussion between a medical professional and data scientists from ORG I about the implication of different intended uses of the future system for the selection of labels. 
\\
\textit{We have two priorities, one is decision support. So it might be easy for you to see the mass, so that won't help you. But there's also the pre-screening - prioritisation. So that might be relevant to detect mass prematurely, right? }(P3)\\
\textit{So if you use it for like a warning, a prioritisation, it can be useful, but for detection... we can see a mass. It's not difficult to find}  (P1).

Medical AI-based systems in our organisations were designed to operate across the world within public or private healthcare systems. Yet medical systems in different countries operate differently based on public values, profit, incentives, and conventions. The design decisions during dataset creation are a product of all these components. The dependency on the healthcare type was well captured by a data scientist from ORG I when discussing the level of detail of the ground truth schema, \textit{"if it was in the US where you actually pay, then from a business point of view, you really wanna find everything. First of all, you don't get sued, and secondly, you can make a lot of money by treating them. But here it's very different, right? Because it's a public system and you only treat things that are necessary, that need to be treated, right?"} (P4). These concerns manifested in debates about what could and needed to be annotated as expert annotators infused the values of their local system into data creation activities.

\subsection{Internal factors: designing the ground truth schema}
While external factors were key in shaping what data was collected and made available for annotation and highlighted the importance of local considerations and their implication for the resulting datasets, two internal factors drove debates, discussions, and disagreements that affected the ground truth schema and the resulting datasets. These were \textbf{Epistemic Differences} and \textbf{Limits of Labelling} (Table \ref{tab:internal}). The effort going into the creation of medical datasets as training data had two purposes that sometimes came into conflict. First, medical datasets were seen as a means of capturing the current state of medical knowledge and the tacit knowledge of medical professionals who focused on medical practice and clinical usefulness. Second, the same datasets served computer scientists as complex input data to solve problems through mathematical operations, where consistency and accuracy were in the spotlight. These two perspectives, while not opposing, often prioritised distinct qualities of the same datasets.

{\small
\begin{table}

\begin{tabular}{p{\linewidth}}
\toprule

\textbf{Epistemic Differences}
\begin{itemize}
    \item Miscommunication between domains
    \item Misapprehension of medical practice
    \item Misapprehension of medical knowledge
\end{itemize}

\textbf{Limits of Labelling}
\begin{itemize}
    \item Domain expert buy-in
    \item Onboarding to the labelling task
    \item Labelling hardware and software
    \item Similarity to the clinical practice
\end{itemize}\\
\bottomrule
\end{tabular}
\caption{Internal factors and their dimensions}
\label{tab:internal}
\end{table}
}

\subsubsection{Epistemic differences}
While in ORG II and ORG III, we engaged with relatively homogeneous teams within each company, in ORG I, our research process was focused on supporting the data creation process by working together with the data science and radiologist teams. As such, in ORG I, we were able to observe first-hand how teams with domain expertise often disagreed on what constituted legitimate knowledge as they discussed what was worth annotating and how things ought to be annotated. We consider three sources of epistemic differences that affected the final design of the ground truth schemas (Table \ref{tab:internal}), communication challenges within the teams, misapprehension of medical practice, and misapprehension of medical knowledge. Within these dimensions, team members from different domains expressed diverging priorities, values, and understanding of concepts, which needed to be reassured and negotiated.

\textbf{Communication challenges within teams.} The three organisations involved stakeholders from different backgrounds, such as health, data science, and design. All of these brought their own traditions, meanings, and domain knowledge that needed to be shared, translated, and understood by other parties for worthwhile collaboration. It is no secret that interdisciplinary teams must spend time finding common ground before they can work together productively \cite{Brown2015Interdisciplinarity:Collaboration}. In our research, we observed how medical professionals, designers, and data scientists constantly translated and explained concepts from their respective domains to maintain a shared understanding. For example, at the beginning of the study in ORG I, medical professionals designed labels based on their, at times naive assumptions of machine learning capabilities, such as when they included two medical concepts under the same label, \textit{"but couldn't that be, if you put nodule, mass in the same category, couldn't you just program it, later on, to say that if the thing that they have marked nodule/mass is over I think ... five millimetres or something, you call it a mass"} (P1), which was not possible given the collected data and was later clarified through a joint discussion. Similarly in ORG II medical professionals had to explain to data scientists that to detect some types of cancer it is necessary to look at more than just the organ in question, and that doctors need to use other information, such as the condition of bile ducts or the blood flow around the organ, affecting data collection and subsequent labelling set up. 

\textbf{Misapprehension of medical practice.} Across the organisations the expectations for the quality of the datasets were closely aligned with concepts such as consistency or bias. This focus was clearly visible when discussing the goal of the labelling task in ORG 1. In the pursuit of consistent and unbiased data, data scientists initially framed labelling as a \textit{"different task"} to clinical work: \textit{"We need to know what's in the image and we need it without them being biased towards looking for only stasis"} (P6). As a result, the labelling task did not provide what was seen by the data scientists as "extraneous and potentially biasing" information, such as the background information of a patient. However, situating the labelling task further away from the medical practice affected the quality of the input medical professionals could provide, impairing the ability of medical professionals to use their knowledge. As one senior radiologist (P10) noted: \textit{"Asking a radiologist to categorise something on a picture only without getting any information on the patient. Is like asking a surgeon to look at the scars on a patient and having him tell you what kind of surgery that patient had"}.

The pursuit of objective and unbiased labels isolated labelling from what data scientists saw as extraneous, potentially biasing information. Yet this transformed the work of the radiologists into a new task that was incompatible with medical practice. To deliver the expected results in this new unfamiliar process, radiologists attempted to reconstruct their medical practice by drawing from their tacit knowledge or, simply, guessing: \textit{I have to create something about the patient myself, which is, [or] might not be true. And I then describe the picture from there...} (P10).

\textbf{Misapprehension of medical knowledge.} Specific data was needed to train AI models that provide clinically useful functionalities. However, due to the misapprehension of practice, the assumptions about what clinical knowledge was possible to extract from the clinical data provided were also at times flawed. As the schema went through iterative rounds of design, we observed how both sides struggled to understand why particular data was requested or why a particular request seemed to be difficult to fulfil. For example, in ORG I, radiologists were asked to assign one of three possible values as a patient's general state based solely on a single chest x-ray, so that relevant cases could be later prioritised using the resulting AI system. This task proved to be particularly problematic to radiologists who do not use such metrics in their daily practice, so they had to develop a range of new approaches to assign them, like \textit{"I chose to interpret it from the view that it could be the worst situation"} (P12) or \textit{"I think it was mostly a gut feeling"} (P11). In the end, the radiologists produced the kind of data that data scientists expected to see as labels. However, what these labels actually captured diverged from the original intention.  

\subsubsection{Limits of labelling}
Finally, we turn to the mechanics of labelling itself that affected the final design of the ground truth schema. We observed schema design and testing in situ directly in ORG I, while in ORG II and ORG III, our data come from post-hoc interviews. We find that four features affected the final design of the ground truth schema (Table \ref{tab:internal}), domain expert buy-in, onboarding to the labelling task, clinical practice familiarity, and labelling hardware and software. These dimensions manifested when evaluating the labelling processes. Unlike the \textit{Epistemic Differences}, where data science was the defining domain, the \textit{Limits of Labelling} emerged as medical professionals confronted the intermediate results of the epistemic negotiations discussed above. These limits altered what kind of data was collected and affected the quality of the labelling.

\textbf{Domain expert buy-in.} Our data showed that domain expert buy-in was crucial and required concessions on the type and amount of collected data. Some ML models require specific types of annotated data, such as \textit{ "what we're asking them is for each patient to go through 500 images and for each image to annotate [...] at pixel level"} (P21). Not only are such tasks typically outside of the scope of clinical practice but are also mentally challenging. For example, when P1 was asked to oversee the labelling process performed by external radiologists, they recalled: \textit{"I think that he [a senior radiologist] opened the program, saw how difficult it was, and just closed it and just never had the energy to start it again"} (P1). Monetary compensation turned out to be a necessary but not sufficient strategy in ORG I for recruiting medical professionals with high expertise to annotate data. 

Once the experts agreed to annotate data, \textbf{limited training for the labelling task reduced the chance for a "shared mindset".} Additional metrics were a relevant part of the ground truth schemas. These metrics usually included concepts not used in daily clinical practice. In ORG I, the medical professionals were supplied with written guidelines to boost common understanding and were briefly introduced to the labelling task. However, some annotators referred to the guidelines only when in doubt: \textit{[the labelling software worked] right out of the box ... I didn't really read this part because it was not necessary} (P12). Not knowing the exact guidelines, medical professionals relied on an intuitive understanding of the metrics and labels, which often resulted in discrepancies between the annotators as they attributed different meanings. 

\textbf{Hardware configuration and user interface of the labelling software affected the quality of the annotations.} These challenges were observed to a greater extent in ORG I, as to assess medical data like CT scans and x-rays, radiologists usually use diagnostic displays. Thus, when they annotate on a \textit{"non-diagnostic screen, you miss details ... maybe small, smaller changes would be missed ... we don't annotate them because we cannot see them" }(P13). Similar comments were shared during the evaluation of the labelling software, medical professionals marked the location of findings using touchpads, which resulted in frustration and low precision. 

Labelling software design could have influenced the final quality of the medical dataset to an even greater extent if not caught during the evaluation. Labelling medical data requires "[a] professional tool that could do the job in a very efficient way” (P21). However, the design of this software could have influenced radiologists in ORG I to overreport radiological findings per x-ray during an evaluation period, \textit{ "...maybe it's an interface. Maybe they forgot the normal button was there because they only saw the [labels]"} (P1). 

The overreporting was not solely caused by the labelling software. \textbf{Expectations and habits influenced what medical professionals noticed in medical data.} For example, a radiologist who reported on an evaluation of the ground truth schema in ORG I reported, \textit{"I told my participants that there would be some normal, but they have not marked any of them normal or I can't find them" }(P1). This phenomenon was later explained by a senior radiologist who pointed to the expectation of labelling a dataset with findings and the fact that when the ratio of abnormal to normal cases is skewed, radiologists tend to overreport to remain on the safe side, \textit{"that's [why] they thought they saw something that was not there"} (P6).

\section{Discussion}
In the creation of high-quality training data, our research shows that the design of ground-truth schema is a crucial but often overlooked step. We highlight five factors that represent external and internal constraints that directly affect the quality of the resulting medical datasets. The external constraints condition the data collection process, affecting this way the design of the ground truth schema, while the internal constraints strongly affect the resulting ground truth schema and can lead to disagreements and debates among domain experts, predominantly data scientists and medical professionals.

\subsection{Conditioning the data collection }
Our findings demonstrate that the regulatory constraints, along with the geographical, demographic, and linguistic context of creation and intended use, and the organisations' scalability crucially affect the amount and type of data that was possible to be collected by the organisations we studied. In this sense, specific data quality metrics were already compromised since the first stage of the medical datasets creation. For example, in ORG I and II the geographical and demographic distribution of the collected data reflected not only how much data was possible to be collected by the contractual agreements in place but also manifested a lack of representativeness, given the regional and local source of data collection.

In ORG III, the aspirations for creating datasets of global coverage stumbled upon the linguistic contextuality of medical terms, which proved to become an issue during the ground truth schema design for the match-making platform. Similarly, in ORG I, the geographical,  demographic, and linguistic context of the medical data collection shaped the type of the collected data, such as that when the experts came to decide on how to design the ground truth schema, dilemmas did not only concern the different understanding of the same medical terms across countries and continents but also possible omissions of local lung diseases. In this sense, the aspiration of designing "transferable" ground truth schemas proved to be both dependent and limited by the standards that regulate the data collection and the context of its collection.

A further insight that emerged in our studies was that the business models and scalability of each organisation affected differently its capacity to collect data. For example, ORG I, being a small-size start-up, having however the public sector involved in its entity, had easier access to timely data (x-ray images of multiple years) from regional hospitals. Yet, the organisation's limited scalability defined the amount of data that was possible to be labelled by medical professionals. In Org III, a similarly small-size start-up, the data collection from both public registries and patients was shaped by the organisation's availability of resources. The constraints were imposed on the recruitment of data scientists designing the platform's ground truth schema and medical professionals who assisted the patients in submitting their medical information into an appropriate and structured format. On the other hand, in ORG II, due to its large size and scalability, the limitations of the data collection were shaped by market demands. This was reflected in the need to collect quality data, i.e., particularly structured, consistent, and contextual medical images from a controlled environment (the contracted local hospital). This push for one type of quality reduced another, in this case, the representativeness of the acquired data.

So far, scholarship has defined and treated data acquisition as a particular step in the data creation process, existing in a vacuum \cite{Nascimento2019UnderstandingSolutions, Hill2016TrialsStudy, Chen2019HowHealthcare, Wang2019Human-AIAI, Amershi2019}. Very little is known about how this step influences the stages that precede the data labelling, eventually affecting the shape of the final medical dataset. Our studies show that regulatory constraints, the context of data creation and use, and the business models and scalability of the organisations, crucially affect the extent and the type of data that is possible to be collected and processed.

\subsection{Conditioning the ground truth design }
Within this context, we identified the design of the ground truth schema as a crucial stage of medical dataset creation. In our studies, the externally imposed constraints shaped the amount and type of data that reached the stage of designing ground truth schema. This has implications for scholarly discussions that focus on developing documentation frameworks that support the responsible and informed use of complex datasets \cite{Anik2021Data-CentricTransparency, Miceli2022DocumentingWork, Gebru2021DatasheetsDatasets, Hutchinson2021TowardsDatasets, Bender2018DataScience}. We showed that the decisions taken during the design of the ground truth schemas were foundational to the succeeding stages of dataset creation. We argue that in this stage, experts do not deal with ideal conditions, but there are inherent limitations which we conceptualised as epistemic differences and limits of labelling. We further argue that the external constraints influence how these inherent limitations manifest in situated collaborative domain settings. 

The amount and type of data that reach the ground truth schema design is already shaped by the necessity of organisations to comply with regulatory standards. This has led the experts from ORG I and II to work with data that had limited representativeness from the start, further affected by the predefined purpose of use and geographical, demographic, and linguistic context for its collection and use. These had implications for the negotiations between data scientists, designers, and medical professionals on what "makes sense" to be labelled. 

Domain negotiations that we observed, were grounded in epistemic differences that did not take place with symmetrically allocated roles, where the “separation of concerns” of each domain expertise is often negotiated against the tacit medical knowledge but where data scientists have the first say \cite{ribes_logic_2019, subramonyam_solving_2022, hutchinson_towards_2021}. Having the development of AI models as the purpose of medical dataset creation, data scientists were positioned as the problem owners of the data creation processes. This further distanced the design of the labels from the medical domain experts and was manifested through misapprehensions about medical knowledge and practice. The tensions with the medical professionals often led to negotiations about what was medically important to be annotated versus what would lead to high-quality datasets from a data science perspective. At the same time, both of these standpoints had to correspond to the demands of the health-tech market.

We found that the externally imposed concerns, such as compliance with regulatory standards, the context of creation, and the intended use of the data, along with the commercial and operational pressures, condition the data collection and can affect ground-truth schema design. In fact, many crucial decisions and negotiations relevant to the final shape of the medical datasets take place during the stage of ground truth schema design. All three organisations under study were committed to developing AI systems in a responsible way. As such, the creation of high-quality training data was a crucial step. Yet, no matter how hard they tried to create representative, consistent, well-structured, high-quality data, the resulting datasets were already limited in different ways. We showed how these limits were predefined even before any data labelling occurred. The combination of external constraints that limit and structure data collection with the misapprehension of domain practice resulted in highly paid experts having to imagine and invent additional information to perform the tasks asked of them. A limited understanding of what is required for diagnosing various conditions from medical images could have  consequences. Either new datasets would have to be created, which translates into a new data collection process, with all the regulatory constraints attached, or the labelling software would have to be more aligned with the existing professional practices following the guidance of expert annotators. Even where these issues were resolved, medical professionals annotated data based on their particular experience and tacit knowledge. This means that the geographical location of the experts affected what they expected to see in the data, showcasing that expertise does not account for the uneven distribution of diseases in different parts of the world.

\section{Limitations and Future Work}
Our contribution builds on qualitative data from three organisations located in countries of the Global North. Creating medical AI datasets in different countries of the Global South may present different challenges and be influenced by a different set of factors that were not captured in our data. Further research is needed to better understand how medical AI data creation varies across different regions and cultures.

Our study focuses on only two medical areas: radiology and clinical trials. While we engaged with diverse types of medical data, creators of other medical datasets could face challenges unique and dependent on different types of medical specialisations. Future research should aim to explore the factors that influence the design of medical AI datasets across a wider range of medical specialisations to develop a more comprehensive understanding of the factors that influence it.

\section{Conclusions}
In this paper, we investigated the work of data scientists, medical professionals, and designers that takes place before the labelling of medical data. Building on the qualitative accounts of our ethnographic findings, our main contributions are:
\begin{itemize}
    \item conceptualising five factors that influence the creation of medical datasets;
    \item disclosing how these factors condition the design of ground truth schemas;
    \item suggesting identified relationships amongst these factors; 
    \item staging the design of the ground truth schemas as a highly contested, yet crucial step in the creation of medical datasets that precedes and conditions data annotation.
\end{itemize}

These overarching factors had a fundamental influence on the final shape of medical datasets created for AI use. First, the externally imposed constraints should be systematically taken into account during the entirety of the medical dataset creation processes, as these factors define the data collection and condition the design of the ground truth schemas. Second, we have exemplified the breadth of decisions taken before the annotation of medical data. Foundational decisions about the final shape of medical datasets take place during the design of a ground truth schema. Future endeavours in data science, law, and policy should consider this stage as crucial to achieving responsible medical AI. 

\begin{acks}
We would like to express our heartfelt gratitude to all of our participants, especially Dr. Elijah Kwasa, Dr. Edward Mwaniki, Dr. Marian Morris, Dr. Ruth Wanjohi, Dr. Mary Onyinkwa, Dr. Sayed Shahnur, and Dr. Samuel Gitau for their invaluable contributions and insightful input. Thank you for taking the time to engage with us and for your significant impact on our work. 
\end{acks}

\bibliographystyle{ACM-Reference-Format}
\bibliography{referencesHubert, references}

\end{document}